\documentclass[journal,10pt,draftclsnofoot,onecolumn]{IEEEtran}

\usepackage{amssymb}
\usepackage{graphics}
\usepackage{graphicx}
\usepackage{epstopdf}
\usepackage{amsmath}
\usepackage{algorithmicx}
\usepackage{algorithm}
\usepackage{multirow}
\usepackage{bm}
\usepackage{enumerate}
\usepackage{array}

\begin{document}
\title{Optimized Compressed Sensing via Incoherent Frames Designed by Convex Optimization}

\author{Cristian Rusu* and Nuria Gonz\'{a}lez-Prelcic \thanks{The authors are with the Atlantic Research Center for Information and Communication Technologies within the University of Vigo, Spain.

E-mail: $\{$crusu,nuria$\}$@gts.uvigo.es. C. Rusu is the corresponding author.}}

\maketitle

\begin{abstract}
The construction of highly incoherent frames, sequences of vectors placed on the unit hyper sphere of a finite dimensional Hilbert space with low correlation between them, has proven very difficult. Algorithms proposed in the past have focused in minimizing the absolute value off-diagonal entries of the Gram matrix of these structures. Recently, a method based on convex optimization that operates directly on the vectors of the frame has been shown to produce promising results. This paper gives a detailed analysis of the optimization problem at the heart of this approach and, based on these insights, proposes a new method that substantially outperforms the initial approach and all current methods in the literature for all types of frames, with low and high redundancy. We give extensive experimental results that show the effectiveness of the proposed method and its application to optimized compressed sensing.
\end{abstract}


\newcounter{remarks}

\section{Introduction}

Frames \cite{BasisAndFrames, LifeBeyondBases1, LifeBeyondBases2} are crucial mathematical objects used for the study of overcomplete bases and subsequent applications. Although they were introduced in the mathematical literature from some time, they have been introduced relatively recently to the signal processing community where their capability to represent signals in overcomplete bases is exploited.

The properties of some classes of frames make them well suited in the area of sparse representations \cite{SparseRep} and compressive sensing \cite{CS}. For the sparse recovery algorithms, like basis pursuit (BP) \cite{BP} and matching pursuit (OMP) \cite{OMP,OMPv2}, a theoretical framework has been developed where the performance of these algorithms can be studied using concepts like the mutual coherence \cite{JustRelax} and the restricted isometry property (RIP) \cite{RIP}. The mutual coherence of a matrix is defined as the maximum absolute value of the normalized dot products between different columns and it is thus very easy to compute, while the RIP characterizes how close a matrix is to an orthogonal one when operating on sparse vectors. The guarantees for successful recovery of the sparsest solutions to overcomplete systems of equations depend directly on these properties, i.e., frames with low mutual coherence are highly desirable. In terms of the mutual coherence the main research direction is to find structures as incoherent as possible (like Equiangular Tight Frames \cite{ETF1}\cite{OnTheExistance}, while for the RIP classes of matrices that meet desirable bounds have been found (we mention random Gaussian and Bernoulli matrices) and used with great success in compressive sensing. Furthermore, even though highly incoherent frames have not been extensively used in signal processing they are of great importance in coding theory and communications \cite{GrassmannianApplication1} for the construction of erasure-robust codes \cite{GoyalErasures,Casazza, Mixon} and for the development of spherical codes \cite{Spherical1} \cite{Spherical2}. Strong connections have been shown to exist between Grassmannian packings, graph theory and spherical codes \cite{GrassmannianApplication2}. Applications have also seen frames used in quantum computing \cite{EldarQuantum}.

Thus, the construction of highly incoherent frames is of central importance.

\subsection{Contribution}

This paper is concerned with the design of real frames as incoherent as possible in an $m$ dimensional Hilbert space with $N$ vectors. We propose an algorithm based on convex optimization called Sequential Iterative Decorrelation by Convex Optimization (SIDCO). We provide insights into the way it behaves, we are able to give conditions under which the algorithm converges, what classes of frames are its minimum and we show experimentally that it substantially outperforms all other methods in the literature. SIDCO is a major improvement of a previously introduced method for incoherent frame design based on convex optimization \cite{IDCO}. With the approach of SIDCO, the problem lends itself to detailed analysis allowing for the selection of the optimal internal parameters, which in \cite{IDCO} are chosen heuristically, while the solutions achieve lower coherence in shorter running times.

The application to optimized compressed sensing \cite{Compare1, Compare2} shows that SIDCO outperforms all previous methods.

\subsection{Previous work}

In the case of frame design, probably the most influential work consists of a design procedure based on the alternating projection method \cite{AlterningProjection} with an implementation presented in \cite{AlterningProjectionCode}. 
These methods concentrate on the Gram matrix of the real frame $\bm{F} \in \mathbb{R}^{m \times N}$ since we need to minimize its largest off-diagonal absolute value entry. The procedures can be summarized by the following iterative procedure: construct the Gram matrix $\bm{G}$ of the current dictionary, decrease the values of a fixed number of entries in $\bm{G}$ by a given amount, use the singular value decomposition (SVD) to reduce the rank of $\bm{G}$ to $m$ and then factor out the new dictionary. Besides the need to run for a large number of iterations, this class of methods proves to work well in practice, especially for small sized dictionaries. Still, there are some internal parameters that need to be tuned: the number of entries of the Gram matrix to be updated and how big this update is. Previous to this family of methods, in the field of communications, algorithms were developed to construct tight frames, that are the optimal sequences for DS-CDMA systems, by minimizing the total squared correlation (TSC) \cite{CDMA3}. Related work on minimizing TSC for binary sequences is also found in \cite{Ding2003}\cite{Ipatov2004}. 
Other methods, like the one in \cite{RoadToEqualNormParsevalFrame} approach a unit norm tight frame by solving a differential equation, while the work in \cite{AutoTuninUnitNormFrames} uses a gradient based method to decrease the coherence of the frame. 

Considerable effort was allocated to the creation of explicit (and sometimes equiangular) frames using constructions by conference matrices \cite{ConfMatrix}, Kerdock codes \cite{Kerdock2}, strongly regular graphs \cite{FramesFromGraphs}, Steiner systems \cite{Steiner} or adjacency matrices of Paley tournaments \cite{Paley}. Although in this paper we study real valued frames, extensive work studies complex \cite{DifferenceSets} and complex harmonic \cite{TroppComplex} frames. For example, in the quantum information theory literature we can find constructions of symmetric, informationally complete, positive operator valued measures (SIC-POVM) \cite{SICPOVM} that achieve low coherence for $N = m^2$ vectors in complex Hilbert spaces. These frames have been constructed, by numerical methods, for all dimensions $m\leq 67$ \cite{SICPOVM_ComputerStudy} and it is conjectured that they exist for all $m$. 

Recently, a method based on averaged projections \cite{Katsagellos} has been proposed where the authors improve the shrinkage operation proposed from \cite{AlterningProjection} to produce less coherent frames. This method is the state of the art in incoherent frame design.

The method that we analyze, and improve upon here, was the first to treat the problem directly on the frame, and not on its Gram matrix. The method \cite{IDCO} solves the original non-convex problem of designing incoherent frames by considering a series of convex problems such that each reduces the coherence of the previous frame.

\subsection{Outline of paper}

The paper is organized as follows: Section \ref{sec:Frames} outlines properties of frames, introduces the optimization problem central to our proposed method and its properties, Section \ref{sec:ProposedMethod} presents the proposed method for incoherent frame design, Section \ref{sec:Results} gives extensive experimental results for the proposed method, with comparisons, and Section \ref{sec:Conclusions} concludes the paper.

\section{Design of incoherent frames}\label{sec:Frames}
In this section we present and analyze the convex optimization problem at the heart of the incoherent frame design method that follows this section.

\subsection{Frames}
We call a finite frame over the field $\mathbb{R}^m$ a sequence of $N$ vectors $\bm{f}_i \in \mathbb{R}^m, 1 \leq i \leq N,$ that satisfies
\begin{equation}
	\alpha \|\bm{v}\|_2^{2} \leq \sum_{i=1}^{N} |\bm{f}_i^T \bm{v}|^2 \leq \beta \|\bm{v}\|_2^{2},\ \forall \ \bm{v} \in \mathbb{R}^m,
	\label{eq:Frames}
\end{equation}
and we introduce the frame synthesis matrix $\bm{F} \in \mathbb{R}^{m \times N}$ consisting of the concatenated vectors of the frame
\begin{equation}
	\bm{F} = \begin{bmatrix} \bm{f}_1 & \bm{f}_2 & \dots & \bm{f}_N \end{bmatrix}.
	\label{eq:FramesF}
\end{equation}
Constants $\alpha$, $\beta \in \mathbb{R}$, with $0 < \alpha \leq \beta < \infty$, are called the lower and upper bounds of the frame. When $\alpha = \beta$ the frame is called $\alpha$-tight and when $\alpha = \beta = 1$ the frame is called a Parseval frame. In this paper, we consider all vectors in the frame to have unit $\ell_2$ norm (i. e. placed on the unit $m$-sphere), which means that the dot products are equal to the cosine of the acute angle between each two vectors. These are called unit norm frames and they are the focus of this paper.

We recall the Gram matrix of the full rank frame $\bm{F}$ to be $\bm{G} = \bm{F}^T \bm{F} \in \mathbb{R}^{N \times N}$. This matrix is symmetric and positive semidefinite with rank $m$, it has a unit diagonal (because of the normalized vectors of the frame) and the off-diagonal entries are equal to the dot products between any two distinct vectors. Based on this matrix we define the mutual coherence as
\begin{equation}
	\mu (\bm{F}) = \underset{1\leq i < j \leq N}{\max}{|g_{ij}|}.
	\label{eq:MutualCoherence}
\end{equation}

From its definition it is clear that the mutual coherence $\mu$ takes values between 0 and 1, with the lower bound reached for orthonormal frames and the upper bound reached when two vectors of the frame are collinear. We define frames that have a low mutual coherence value to be incoherent. The mutual coherence is useful, for example, in the context of sparse representations since it provides a bound on the performance of sparse approximation algorithms \cite{TheSparseBound}: considering the system $\bm{F}\bm{x}=\bm{y}$, with $\bm{F}$ full rank $m$, the solution $\bm{x}$ of BP or OMP is the sparsest if
\begin{equation}
	\|\bm{x}\|_0 \leq s =  \frac{1}{2} \left( \frac{1}{\mu (\bm{F})} + 1 \right).
	\label{eq:mcResult}
\end{equation}

The frame potential \cite{FiniteNormalizedTightFrames}
\begin{equation}
	\text{FP}(\bm{F}) = \sum_{i = 1}^N \sum_{j=1}^{N} | g_{ij} |^2 = \| \bm{G} \|_F^2,
	\label{eq:FP}
\end{equation}
was introduced to describe how close the frame $\bm{F}$ is to a unitary tight frame, when FP achieves its minimum of $N^2/m$.

Since the properties of interest of these frames revolve around the Gram matrix, most design procedures consider the $\bm{G}$ matrix the central object while their constructions are based on the following theorem presented in \cite{AlterningProjection} and the numerical procedure of polar decomposition to recover the frame.

\textbf{Theorem 2 of \cite{AlterningProjection}.} Let $\bm{F}$ be a $m \times N$ matrix with singular value decomposition $\bm{U\Sigma V}^H$. With respect to the Frobenius norm, the closest $\alpha$-tight frame to $\bm{F}$ is given by $\alpha\bm{UV}^H$. \hfill$\blacksquare$

We next define the Equiangular Tight Frames (ETFs) \cite{ETF1} as the sequence of $N$ vectors $\bm{f}_j$ that obey:
	\begin{equation}
		\|\bm{f}_i \|_2^2 = 1 \text{ and } |\bm{f}_i^T \bm{f}_j | = \mu,\ 1 \leq i \leq j \leq N,i \neq j
		\label{eq:ETFProp1and2}
	\end{equation}
	\begin{equation}
		\frac{m}{N} \sum_{i=1}^{N} ( \bm{f}_i^T \bm{v})\bm{f}_i = \bm{v}, \forall \ \bm{v} \in \mathbb{R}^{m},
		\label{eq:ETFProp3}
	\end{equation}
with an immediate consequence of \eqref{eq:ETFProp1and2} and \eqref{eq:ETFProp3} that the distances between all the different member vectors are the same and equal, in absolute value, to
\begin{equation}
	\mu = \sqrt{ \frac{ N-m }{m (N-1)}}.
	\label{eq:FrameBound}
\end{equation}
This relation describes in fact a general lower bound on the mutual coherence that is also known as the Welch bound (WB) or Ranking bound \cite{WB, Waldron} which we use as a reference point performance, for perspective. This is the optimal bound of coherence for $N \leq m^2$, the regime that we consider here.

The ratio $\rho = N/m$ defines the redundancy of the frame and we call frames $(m,km)$ to be $k$--overcomplete. 

It is trivial to observe that orthonormal frames are ETFs and actually reach the lowest possible coherence, zero. Sadly, we are interested in highly overcomplete frames, i.e., $N \gg m$, where ETFs only exist for a few pairs $(m,N)$ (for example, for the real case, these do not exist if $N > m(m+1)/2$ \cite{OnTheExistance}). When they exist, ETFs provide the lowest possible mutual coherence and they are considered the generalization of orthonormal bases because they preserve (a generalization of) Parseval's identity \eqref{eq:ETFProp3}.

A generalization of ETFs is embodied by Grassmannian Frames (GFs). For these structures, the largest absolute value dot product is minimum for the given dimensions $m$ and $N$. Notice that this relaxed condition means that GFs exist for every pair $(m,N)$. Still, this does not help in decreasing the computational complexity of finding them or even deciding if a given frame is a GF.

From this brief outline it should be clear that creating maximally incoherent frames (either ETF or Grassmanian) is an extremely difficult task in the general case and there are no guarantees that the lower bound of coherence can be reached for a given pair $(m,N)$. Necessary conditions for the existence of ETFs are given in \cite{OnTheExistance, Holmes}.

\subsection{Incoherent frame design}

This subsection is dedicated to the description of a numerical procedure for the design of incoherent frames. We develop an optimization procedure that directly designs the frame $\bm{F}$.

Ideally, we would like to solve exactly the following non-convex optimization problem:
		\begin{equation}
			\begin{aligned}
			& \underset{\bm{F}}{\text{minimize}} & & \underset{i,j;\ j \neq i}{\max \ }{|\bm{f}_j^T \bm{f}_i|} \\
			& \text{subject to} & & \| \bm{f}_i \|_2^2 = 1, \ 1 \leq i \leq N.
			\label{eq:IDCOMain}
			\end{aligned}
		\end{equation}
Solving this problem exactly would produce ETFs, when they exist, or Grassmanian frames otherwise. The difficulty lies in the fact that both the objective and the equality constraints are non-convex. We proceed to relax this problem and provide convex optimization formulations to solve it approximately.

\section{The proposed algorithm}\label{sec:ProposedMethod}

The approach taken in this paper is not to solve \eqref{eq:IDCOMain} directly but given a frame $\bm{H} = \begin{bmatrix} \bm{h}_1 & \dots & \bm{h}_N \end{bmatrix} \in \mathbb{R}^{m \times N}$ to find a new frame $\bm{F} = \begin{bmatrix} \bm{f}_1 & \dots & \bm{f}_N \end{bmatrix}$ that is near (in a sense that is made exact) to the initial one, with smaller mutual coherence.

For all $i = 1,\dots,N$ the optimization problem we study is
		\begin{equation}
			\underset{\bm{f}_i; \ \|\bm{f}_i - \bm{h}_i\|_2^2 \leq T_i}{\text{minimize}} \ \  \underset{j;\ j \neq i}{\max}\ |\bm{h}_j^T\bm{f}_i|.
			\label{eq:IDCOMain2}
		\end{equation}
For clarity, we can also define $\bm{H}_i = \begin{bmatrix} \bm{h}_1 & \dots & \bm{h}_{i-1} & \bm{h}_{i+1} & \dots & \bm{h}_N \end{bmatrix}$ and thus the objective function of \eqref{eq:IDCOMain2} becomes $\|\bm{H}_i^T\bm{f}_i\|_\infty$. Solving this problem, we define for each vector $\bm{h}_i$ in the reference frame a trust region (a closed $m$-ball centered at $\bm{h}_i$ of radius $\sqrt{T_i}$) where we search for a new vector $\bm{f}_i$ such that its correlation (in absolute value) with the other vectors in the frame $\bm{H}_i$ is smaller than that of $\bm{h}_i$. The only input parameter of the optimization procedure $T_i \in \mathbb{R},\ 0 < T_i \ll 1,$ is a measure of how far we are willing to leave the reference frame $\bm{H}$ in the search for the new frame $\bm{F}$. Since now we are dealing with a convex objective function and constraint, the formulation in \eqref{eq:IDCOMain2} can be viewed as the convex relaxation of \eqref{eq:IDCOMain}.

Observe that each vector of the frame can be computed separately in \eqref{eq:IDCOMain2}. Given a fixed frame $\bm{H}$ assume that $\bm{f}_i$ is the only optimization variable and, without loss of generality, that $\bm{h}_j^T \bm{h}_i \geq 0, 1 \leq j \leq N$ (when not, flip the sign of $\bm{h}_j$). We stress this form of the frame since this imposed structure reduces the size of the convex optimization problems that need to be solved and, as we will see in the following subsections, highlights several properties of frames. 

Problem \eqref{eq:IDCOMain2} can be separated into $N$ convex optimization problems for $i=1,\dots,N$
		\begin{equation}
			\underset{\bm{f}_i; \ \|\bm{f}_i - \bm{h}_i\|_2^2 \leq T_i}{\text{minimize}} \ \  \underset{j;\ j \neq i}{\max}\ (\bm{h}_j^T\bm{f}_i)
			\label{eq:IDCOMain3}
		\end{equation}
This optimization problem lies at the heart of the proposed incoherent frame design method. Since the solution will lead to vectors that are not of unit norm (due to the inequality constraint) a normalization step must always follow. Solving \eqref{eq:IDCOMain3} can be done in polynomial time using, for example, optimization strategies like interior point methods \cite{CO}.

Notice that for a given frame, the distances between its vectors (and consequently its mutual coherence) are preserved under column permutations and orthogonal (rotational) transformations. Given a frame $\bm{H}$ with Gram matrix $\bm{G} = \bm{H}^T \bm{H}$, if we rotate the frame using an orthonormal transformation $\bm{Q} \in \mathbb{R}^{m \times m},$ i.e., $\bm{Q}^T \bm{Q} = \bm{I},$ then the new frame $\bm{H}_\text{new} = \bm{QH}$ has the same Gram matrix since $\bm{G}_\text{new} = \bm{H}_\text{new}^T \bm{H}_\text{new} = \bm{H}^T\bm{Q}^T \bm{Q} \bm{H} = \bm{H}^T \bm{H} = \bm{G}$. Following this observation notice that we can fix the first vector of a frame to any arbitrary vector (e.g., $\bm{e}_1$) since we can always compute a rotation $\bm{Q}$ such that the first vector has a predefined, fixed, position. Thus, we can solve problem \eqref{eq:IDCOMain3} for $i=2,\dots,N$.

All these properties will be used to simplify the structure of the proposed algorithm. We next proceed to study some of the properties of \eqref{eq:IDCOMain3}.

\subsection{The choice of parameter $T_i$}

In this section we discuss the choice of the only parameter needed $T_i$. Since the original equality constraint $\| \bm{f}_i \|_2^2 = 1$ is not tractable in \eqref{eq:IDCOMain}, we analyze the consequences of the relaxed inequality constraint introduced in \eqref{eq:IDCOMain3}, $\| \bm{f}_i - \bm{h}_i \|_2^2 \leq T_i$.

\begin{figure}[t]
\centering
\includegraphics[width=0.45\textwidth]{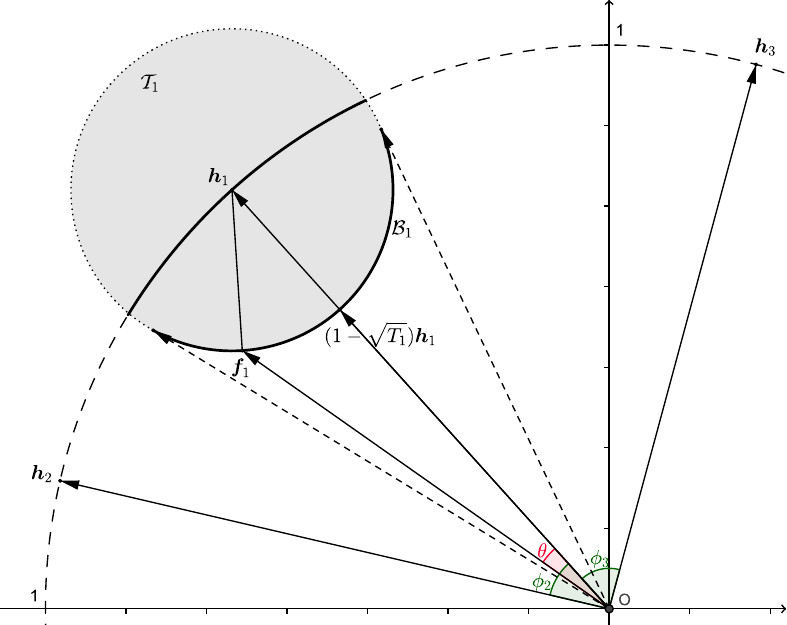}
\caption{A 2D graphical description of the trust region $\mathcal{T}_1$, hashed in grey, centered at $\bm{h}_1$ with fixed radius $\sqrt{T_1}$ and the boundary $\mathcal{B}_1$ depicted by the bold arc of $\mathcal{T}_1$ where all solutions lie in a typical situation generated by $\| \bm{f}_1 - \bm{h}_1 \|_2^2 = T_1$.}
\label{fig:ChoiceOfT}
\end{figure}

In the optimization problem \eqref{eq:IDCOMain3} we define for each vector belonging to the reference frame $\bm{H}$ a trust region, called here $\mathcal{T}_i$, represented by a closed $m$-ball centered around the reference vector $\bm{h}_i$ of fixed radius $\sqrt{T_i}$. It is in this trust region that we search for a new vector $\bm{f}_i$ that replaces the reference $\bm{h}_i$ such that the coherence of the frame is reduced. The question is how to choose $T_i$ such that we can guarantee that the mutual coherence of the frame is not increased after the normalization of the solution to \eqref{eq:IDCOMain3}. First, notice that the solution to \eqref{eq:IDCOMain3} will always be on the hyperspherical cap, denoted here by $\mathcal{B}_i$, of the trust region $\mathcal{T}_i$ that is bounded by the tangents to $\mathcal{T}_i$ from the origin. This is true because any solution $\bm{f}_i \in \mathcal{T}_i$ produces a dot product higher than its projection on $\mathcal{B}_i$ -- a scaled $\gamma \bm{f}_i \in \mathcal{B}_i, 0 < \gamma \leq 1$. This leads to the conclusion that the inequality constraint is satisfied always with equality $\| \bm{f}_i - \bm{h}_i \|_2^2 = T_i$. Since $T_i>0$, the solution $\bm{f}_i$ is never on the unit $m$-sphere and thus a normalization step must follow -- this is due to the fact that $\{\bm{f}_i \ | \ \bm{f}_i \in \mathcal{B}_i\} \cap  \{\bm{f}_i \ | \ \bm{f}_i \in \mathcal{T}_i \text{ and } \|\bm{f}_i\|_2^2 =1\} = \emptyset$. A graphical description of these regions is shown in Figure \ref{fig:ChoiceOfT}.

When there is no change in the direction of the new vector of the frame the solution is $\bm{f}_i = (1-\sqrt{T_i})\bm{h}_i$. Otherwise, a change $\theta$ in angle from $\bm{h}_i$ might cause a reduction in coherence with the other vectors of the frame. Because the solution belongs to the cap $\mathcal{B}_i$, any deviation $\theta$ from the reference vector leads to a change in the norm of the solution
\begin{equation}
	\| \bm{f}_i \|_2(\theta) = \cos\theta - \sqrt{\cos^2\theta - 1 + T_i},
	\label{eq:lengthoffi}
\end{equation}
for $0 \leq \theta \leq \arccos\sqrt{1-T_i}$. This leads to $1 - \sqrt{T_i} \leq \| \bm{f}_i \|_2(\theta) \leq \sqrt{1-T_i}$. Relation \eqref{eq:lengthoffi} follows by the law of cosines in the triangle with vertices in the origin, $\bm{h}_i$ and any point $\bm{f}_i$ in the region $\mathcal{B}_i$: $T_i = 1 + \|\bm{f}_i\|_2^2 - 2\|\bm{f}_i\|_2^2\cos\theta$.

\stepcounter{remarks}
\textbf{Remark \arabic{remarks}. (The choice of $\bm{T_i}$ and the convergence of the proposed algorithm):} The solution region $\mathcal{B}_i$ must not contain any vector collinear with the other vectors $\bm{h}_j, j \neq i$. For vector $\bm{h}_i$ if the smallest angle with the other vectors is denoted by $\phi^\text{min}$ then we need to have $\theta^\text{max} < \phi^\text{min}$ which reads $\arccos\sqrt{1-T_i} < \phi^\text{min}$ and finally develops in
\begin{equation}
	T_i < 1-\cos^2 \phi^\text{min} = 1 - \underset{j;\ j \neq i}{\max}\ |g_{ij}|^2.
	\label{eq:chooseT}
\end{equation}
Consider that after solving \eqref{eq:IDCOMain3} for a fixed $i$ with the $T_i$ computed by \eqref{eq:chooseT} we obtain the solution $\bm{f}_i$ which makes an angle $\theta$ with $(1-\sqrt{T_i})\bm{h}_i$. This means that
\begin{equation}
	\| \bm{H}_i^T \bm{f}_i \|_\infty \leq \| \bm{H}_i^T \bm{h}_i \|_\infty (1-\sqrt{T_i}).
	\label{eq:result123}
\end{equation}
To finalize the new frame vector the normalization step that places the solution $\bm{f}_i$ on the unit $m$--sphere follows. We would like the normalized solution $\bm{f}_i( \| \bm{f}_i\|_2(\theta))^{-1}$ to still be as good as (or better than) $\bm{h}_i$, that is
\begin{equation}
	\| \bm{H}_i^T \bm{f}_i \|_\infty  \leq \| \bm{H}_i^T \bm{h}_i \|_\infty \| \bm{f}_i \|_2 (\theta),
\end{equation}
This inequality is always true since \eqref{eq:result123} is true and $\| \bm{f}_i \|_2 (\theta) \geq 1-\sqrt{T_i}$. Intuitively this makes sense since the norm of the solution $\bm{f}_i$ of \eqref{eq:IDCOMain3} increases with the increase in the angle $\theta$, by \eqref{eq:lengthoffi}. Ultimately, this means that the solution $\bm{f}_i$ provides a coherence level at least as small as $\bm{h}_i$ and therefore no increase in coherence can happen. The proposed algorithm solves problem \eqref{eq:IDCOMain3} for each vector of the frame and thus the mutual coherence is monotonically decreasing with each step solving \eqref{eq:IDCOMain3} with parameters given by \eqref{eq:chooseT}. Since the coherence is bounded below, the convergence of the overall algorithm to a local minimum is guaranteed.
\hfill$\blacksquare$

\stepcounter{remarks}
\textbf{Remark \arabic{remarks}. (ETFs are local minimum points):} Given an ETF, any solution of \eqref{eq:IDCOMain3} for $\bm{h}_i$ that is different from $\bm{f}_i =(1-\sqrt{T_i}) \bm{h}_i$ causes an increase in the correlations with the other vectors not only because of the decrease in the angle between at least one of them but also since any deviation $\theta$ from $\bm{h}_i$ causes a strict increase in \eqref{eq:lengthoffi}. Thus, ETFs are local minimum points of \eqref{eq:IDCOMain3} (up to the normalization factor). \hfill$\blacksquare$

The results presented in Remark 1 show the effect that parameter $T_i$ has in the optimization problem and guidelines on how to choose it in order to ensure and maximize the decrease in coherence. Next we deal with the overall optimization problem.

\subsection{Analysis of the optimization problem}

In this section we analyze the optimization problem at the heart of the proposed method. Since in the previous section we detail the effect of the parameters $T_i$, we are now mainly concerned with the convex relaxation made in the objective function. In this section we analyze optimization problem \eqref{eq:IDCOMain3}:
		\begin{equation*}
			\underset{\bm{f}_i; \ \| \bm{f}_i - \bm{h}_i \|_2^2 \leq T_i}{\text{minimize}} \ \  \underset{j;\ j \neq i}{\max}\ (\bm{h}_j^T\bm{f}_i),
		\end{equation*}
which written in a standard form reads
		\begin{equation}
			\begin{aligned}
			& \underset{\bm{x} = [\bm{f}_i; \ t]}{\text{minimize}} & & \bm{c}^T \bm{x} \\
			& \text{subject to} & & \bm{Ax} \leq \bm{0} \\
			& & & \bm{x}^T  \bm{Bx} - 2\bm{b}^T \bm{x} + 1 - T_i \leq 0,
			\label{eq:IDCOMain5}
			\end{aligned}
		\end{equation}
where $\bm{B} = \begin{bmatrix} \bm{I} & \bm{0} \\ \bm{0}^T & 0 \end{bmatrix} \in \mathbb{R}^{(m+1)\times (m+1)}, \bm{b}^T = \begin{bmatrix} \bm{h}_i^T & 0 \end{bmatrix} \in \mathbb{R}^{m+1}, \bm{c}^T = \begin{bmatrix} \bm{0}_{1 \times m}^T & 1 \end{bmatrix} \in \mathbb{R}^{m+1}, \ \bm{A} = \begin{bmatrix} \bm{H}_i^T & -\bm{1}_{(N-1) \times 1} \end{bmatrix} \in \mathbb{R}^{(N-1) \times (m+1)}$ and $t \in \mathbb{R}_+$ holds the $\ell_\infty$ objective function value. For the solution $\bm{x} \in \mathbb{R}^{m+1}$ the Karush-Kuhn-Tucker (KKT) conditions read:
\begin{equation}
	\begin{array}{ll}
	\bm{Ax} \leq \bm{0}, & \bm{x}^T  \bm{Bx} - 2\bm{b}^T \bm{x} + 1 - T_i\leq 0, \\
	\lambda_j (\bm{Ax})_j = 0,  & \lambda_N (\bm{x}^T  \bm{Bx} - 2\bm{b}^T \bm{x} + 1-T_i) = 0, \\
	& \bm{c} + \bm{A}^T \bm{\lambda} + 2 \lambda_N ( \bm{Bx} - \bm{b}) = \bm{0},
	\end{array}
\end{equation}
with $j=1,\dots,N-1$ and where $\bm{\lambda} \in \mathbb{R}^{N-1}_+$, $\lambda_N \in \mathbb{R}_+$ are the Lagrange multipliers. Develop the last equality to get
\begin{equation}
	\left\{ \begin{aligned}
	& \bm{H}_i \bm{\lambda} = -2 \lambda_N (\bm{f}_i - \bm{h}_i), \\
	& \sum_{j=1}^{N-1} \lambda_j = 1 \text{ and } \lambda_j = 0 \text{ when } \bm{h}_j^T \bm{f}_i \neq t.
	\end{aligned} \right.
	\label{eq:basicrelation}
\end{equation}
Multiply both side by $\bm{f}_i^T$ and notice that $\bm{f}_i^T \bm{H}_i\bm{\lambda} = t$ to reach
\begin{equation}
         t = 2 \lambda_N (\bm{f}_i^T\bm{h}_i - \| \bm{f}_i \|_2^2).
	\label{eq:t}
\end{equation}
Since $\lambda_N \geq 0$ and, by \eqref{eq:lengthoffi}, $\bm{f}_i^T\bm{h}_i - \| \bm{f}_i \|_2^2 = \| \bm{f}_i\|_2\cos\theta -  \| \bm{f}_i\|_2^2 =  \| \bm{f}_i\|_2 \sqrt{\cos^2\theta - 1 + T_i} \geq 0$ we have that $t\geq 0$. Notice that $\lambda_N$ may be nonzero since the associated inequality constraint $ \| \bm{f}_i - \bm{h}_i \|_2^2 \leq T_i$ is always met with equality (by the argument provided in the previous section).\hfill $\blacksquare$

\stepcounter{remarks}
\textbf{Remark \arabic{remarks}. (On the existence of real equiangular tight frames):} Given an ETF $\bm{H} \in \mathbb{R}^{m \times N}$ define the frame $\bm{H}_i = \begin{bmatrix} \bm{h}_1 & \dots & \bm{h}_{i-1} & \bm{h}_{i+1} & \dots & \bm{h}_N \end{bmatrix} \in \mathbb{R}^{m \times (N-1)}$, assuming that for fixed $i$ we have $\bm{h}_i^T \bm{h}_{j} = \mu, j \neq i$, then $\bm{H}_i^T \bm{H}_i$ has eigenvector $\bm{1} \in \mathbb{R}^{N-1}$ with eigenvalue $(N-m)/m$.

To see this develop \eqref{eq:ETFProp3} for $\bm{v} = \bm{h}_i$ to get
\begin{equation*}
		\bm{h}_i = \frac{m}{N} \sum_{j=1}^N (\bm{h}_i^T \bm{h}_j) \bm{h}_j = \frac{m}{N} \left( \sum_{ \underset{j\neq i}{j=1} }^N \mu \bm{h}_j  + \bm{h}_i \right)
				= \frac{m \mu}{N-m} \bm{H}_i \bm{1},
\end{equation*}
and multiply both sides by $\bm{H}_i^T$ to finalize
\begin{equation}
		\frac{m \mu}{N-m} \bm{H}_i^T \bm{H}_i \bm{1} = \bm{H}_i^T \bm{h}_i  \Rightarrow
		 (\bm{H}_i^T \bm{H}_i) \bm{1} = \frac{N-m}{m} \bm{1}.
		\label{eq:eigenresult}
\end{equation}

We use these results to show how the system of equations in \eqref{eq:basicrelation} behaves when dealing with an ETF. Also, result \eqref{eq:eigenresult} has several direct consequences for $\bm{H}$:
\begin{enumerate}
	\item With the assumptions made on $\bm{h}_i$ the Gram matrix is
\begin{equation}
	\bm{G} = \bm{H}^T\bm{H} = \begin{bmatrix}
				\star & \mu \bm{1} & \star \\
				\mu \bm{1}^T & 1 & \mu \bm{1}^T\\
				\star & \mu \bm{1} & \star 
			\end{bmatrix}.
	\label{eq:newGram}
\end{equation}

	\item Each vector of the frame $\bm{h}_j$ has the same number of positive ($N^+ _j$) and negative ($N^- _j$) correlations with the other vectors in the frame, except for the $i^{\text{th}}$ vector for which $N^+_i = N-1$ and $N^-_i = 0$.

	\item With \eqref{eq:newGram}, equality in \eqref{eq:eigenresult} and the additional positive correlation with $\bm{h}_i$ we obtain $(N^+ _j -1 - N^- _j) \mu + 1= (N-m)/m$ for all $\bm{h}_j, j \neq i$. Use \eqref{eq:FrameBound} and we have
	\begin{equation}
		N^+ _j - N^- _j = \frac{N - 2m}{m \mu} +1.
	\label{eq:Holmes}
	\end{equation}
The fact that the first term on the right hand side is an integer for ETFs was already shown in \cite{Holmes}. This result offers another possible interpretation of that quantity. Information about the signs in the Gram matrix can be used to compute, for example, the average coherence \cite{AverageCoherence} for ETFs. This concept was linked to the success of sparse approximation algorithms. $\hfill \blacksquare$
\end{enumerate}

\stepcounter{remarks}
\textbf{Remark \arabic{remarks}. (The KKT conditions for ETFs):} Considering Remark 2, when dealing  with an ETF, we have that $\bm{f}_i = (1-\sqrt{T_i})\bm{h}_i$, $t = (1-\sqrt{T_i})\mu$ and thus \eqref{eq:t} reduces to $\lambda_N = \mu(2\sqrt{T_i})^{-1}$. As a consequence in \eqref{eq:basicrelation} we have that
\begin{equation}
	\bm{H}_i\bm{\lambda} = \mu \bm{h}_i. 
	\label{eq:tForETF}
\end{equation}
With this result and \eqref{eq:eigenresult} observe that
\begin{equation}
	\bm{h}_i = \frac{m \mu}{N-m} \bm{H}_i\bm{1} = \frac{1}{\mu}\bm{H}_i\bm{\lambda},
\end{equation}
which leads to $\bm{\lambda} = \frac{m \mu^2}{N-m}\bm{1} = \frac{1}{N-1}\bm{1} \in \mathbb{R}_{+}^{N-1}. \hfill\blacksquare$

\stepcounter{remarks}
\textbf{Remark \arabic{remarks}. (Description of some local minimum points):} Let $J$ denote the set of indices $j$, with $j \neq i$, for which $\bm{f}_i^T\bm{h}_j = t$ and let $\bm{H}_J$ be the matrix $\bm{H}$ constraint to the columns indexed by $J$. From \eqref{eq:basicrelation} it is clear that the change $\bm{r}_i = \bm{h}_i - \bm{f}_i$ is in the column space of $\bm{H}_J$.

Consider now a frame for which $\bm{H}_J^T \bm{h}_i = \mu_1\bm{1}$, where $\mu_1$ has the mutual coherence value, and without loss of generality that $J = \{1,\dots,m\}$, and thus $\bm{H}_J$ is of full rank and has empty null space $\mathcal{N}(\bm{H}_J) = \emptyset$ (we do not consider $\bm{0}_{m \times 1})$. Then consider that we are looking for a direction $\bm{r}_i$ such that $\bm{H}_J^T (\bm{h}_i-\bm{r}_i) = \mu_2 \bm{1}$ and additionally $\bm{h}_{m+1}^T\bm{h}_i = \mu_2$ with $\mu_2 \leq \mu_1$ -- in effect we are looking to find a direction such that the coherence is maximal with one extra frame vector. With the given full rank $\bm{H}_J$ the only possible solution is thus $\bm{r}_i = (\mu_1 - \mu_2)/\mu_1 \bm{h}_i$. This means that the algorithm cannot make any progress. Of course the results holds when $\bm{H}$ is an ETF and then $|J| = N-1, \mu_1 = \mu, \mu_2 = (1-\sqrt{T_i})\mu$ and thus $\bm{r}_i = \sqrt{T_i}\bm{h}_i$.

Since we are also looking for highly overcomplete frames ($N \gg m$) this result might be worrying. To overcome this drawback, we will introduce a heuristic step in the proposed method that deals with such situations and allows the optimization problems to make further progress after getting stuck in such local minimum points. \hfill$\blacksquare$

\stepcounter{remarks}
\textbf{Remark \arabic{remarks}. (On the constraint that defines the trust region $\bm{\mathcal{T}_i}$):} In this implementation of \eqref{eq:IDCOMain3} we use an $\ell_2$ constraint to define the trust region around each reference frame vector. We could use other trust regions, like the ones defined by the $\ell_1$ or $\ell_\infty$ norm. These "box" constraints have the extra benefit of keeping the optimization problem \eqref{eq:IDCOMain2} to a linear program (as defined in \eqref{eq:IDCOMain3}, the objective is already linear). Of course, since the definition of the trust region is crucial to the choices of $T_i$ and the analysis of the algorithm, the results presented in this and the previous sections, does not hold anymore. \hfill$\blacksquare$

\stepcounter{remarks}
\textbf{Remark \arabic{remarks}. (Reducing the complexity of the optimization problem):} We have seen in Remark 5 that $\bm{r}_i$ is in the column space of $\bm{H}_J$ where $J$ is the set of indices that achieve maximum coherence with $\bm{f}_i$. For some $j$, it may be that the correlation of $\bm{h}_j$ with $\bm{h}_i$ is so low that given the trust region constraint there is no way this index $j$ will belong to the set $J$. We want to solve \eqref{eq:IDCOMain3} only for indices $j, j \neq i,$ that may participate in $J$. Formally, we want to remove from the optimization problem, from $\bm{H}_i$, the vectors $j$ for which
\begin{equation}
	\cos(\phi_j - \theta^\text{max}) < \cos(\phi^\text{min} + \theta^\text{max}),
	\label{eq:whattodrop}
\end{equation}
where the left hand side is the maximum coherence that can be achieved by vector $\bm{h}_j$ with $\bm{f}_i$, after normalization, and the right hand term describes the maximum decrease in mutual coherence that can be achieved, both after solving \eqref{eq:IDCOMain3} ($\phi_j$ is the angle between $\bm{h}_i$ and $\bm{h}_j, j \neq i$, $\phi^\text{min} = \min_j \phi_j$ and $\theta^\text{max}$ is the maximum allowed angle change from the reference $\bm{h}_i$). Together with equation \eqref{eq:chooseT} that states $\theta^\text{max} = \phi^\text{min}$, \eqref{eq:whattodrop} reduces to $\phi_j > 3 \phi^\text{min}$. This step proves efficient when the frame is coherent) -- for an ETF nothing is dropped. \hfill$\blacksquare$

The results and the remarks presented in this section come together to construct an efficient method for incoherent frame designed which is described in the following section.

The proposed method, called Sequential Iterative Decorrelation by Convex Optimization (SIDCO) gradually decreases the coherence of an initial frame by finding a new frame that is close to the previous one and achieves lower coherence.

The general structure of SIDCO is represented by the following iterative optimization procedure that updates each vector of the frame in the sequence one at a time. The method brings together the results presented in the previous sections.

\begin{algorithm}[t]

\caption{ \textbf{Sequential Iterative Decorrelation by Convex Optimization (SIDCO)}. Given dimensions ($m$, $N$), construct frame $\bm{H} \in \mathbb{R}^{m \times N}$ as incoherent as possible until maximum number of iterations $K$ is reached. }

\begin{algorithmic}

\State \textbf{Initialization}:
\begin{enumerate}

	\item Create $\bm{H} \in \mathbb{R}^{m \times N}$ with random entries from a Gaussian distribution. Normalize its columns.

	\item With $\bm{H} = \bm{U \Sigma V}^T$ update the frame by the unit polar decomposition (by Theorem 2 of \cite{AlterningProjection}): $\bm{H} = \bm{U}\bm{V}^T$. 

	\item Normalize the columns of $\bm{H}$. Set reference frame $\bm{H}$.
\end{enumerate}

\State \textbf{Iterations} $k = 1,\dots ,K$:
	\begin{enumerate}

		\item For each $i$ in randomized $\{ 2,\dots,N \}$:
		\begin{enumerate}
			\item Flip the sign of $\bm{h}_j$ where $\bm{h}_i^T \bm{h}_j < 0$ for all $j \in \{1,\dots,N\} \backslash \{i\}$.

			\item Compute radius $T_i$ of trust region $\mathcal{T}_i$ by \eqref{eq:chooseT}.

			\item Define $J = \{ j\ | \text{ such that } \bm{h}_i^T \bm{h}_j \text{ is maximum}\}$. \newline If $|J| < m$
reduce problem dimension by \eqref{eq:whattodrop} and solve optimization problem \eqref{eq:IDCOMain3} for $\bm{f}_i$.
			\item Normalize and update $\bm{h}_i = \bm{f}_i \| \bm{f}_i \|_2^{-1}$.

		\end{enumerate}

		\item If the algorithm has converged, update frame by its unit polar decomposition $\bm{H} = \bm{UV}^T$ (again by Theorem 2 of \cite{AlterningProjection}) and normalize columns.

	\end{enumerate}

\end{algorithmic}
\end{algorithm}
Again, we stress that this optimization procedure works by directly modifying the frame and not its Gram matrix. The SIDCO approach is somewhat similar to the sequential convex programming framework that approximates solutions to nonconvex optimization problems by solving a series of locally convex optimization problems in a defined trust region.

\textbf{The initialization.} The initial frame we consider has its entries drawn randomly from a Gaussian distribution. This simple initialization is useful in this case since it has been shown that random frames have very low coherence values and actually, as $N$ tends to infinity, they converge to tight frames \cite{GV}. Vectors from a Gaussian distribution are uniformly placed on the unit $m$-sphere since the Gaussian distribution is rotational invariant. By Theorem 2 of \cite{AlterningProjection}, we next perform a polar decomposition of the frame and we reconstruct using only the unit part of the decomposition. To finalize, normalize the frame vectors. Herein, the initial reference frame is $\bm{H}_0$.

\textbf{The iterations.} By the argument provided in the previous sections, at each iteration of SIDCO we solve a sequence of $N-1$ convex optimization problems that produce the vectors of a new frame such that the mutual coherence is decreased. The frame vectors are chosen in a randomized way so to further avoid the convergence in a local minimum point. Before solving the optimization problem, several choices need to be made. A discussion of these follows.

First, we change the orientation of the frame vectors such that the current variable vector has positive correlation with all the other fixed vectors. This allows for both a smaller optimization problem and a simplified analysis of the optimization problem \eqref{eq:IDCOMain3}. We pick the maximum size of the trust region $T_i$ which also guarantees that the mutual coherence of the frame will not increase after solving the optimization problem \eqref{eq:IDCOMain3}, as discussed in Remark 1. We check if the current frame vector $\bm{h}_i$ can be improved and if so we reduce the dimension of the optimization problem, by the arguments provided in Remark 5 and Remark 7, respectively. We finalize the new vector by a normalization step placing it on the unit $m$-sphere.

Since Remark 5 states that the algorithm does not update a frame vector when it reaches the maximal correlation with a number of $m$ other frame vectors, we add a heuristic step to push the current frame from its local minimum if the algorithm has converged too quickly. We say that the algorithm has converged if in the last 3 iterations the average progress in mutual coherence is less than $\epsilon_\text{stop} = 10^{-5}$. The step applies a polar decomposition as described by Theorem 2 of \cite{AlterningProjection}. We expect that this update will lead to a relatively small temporary increase in the mutual coherence that will be reduced in the following regular iterations, hopefully to levels lower than the one before the application of this step. As we will show experimentally, SIDCO makes significant progress in the reduction of the mutual coherence in a relative small number of iterations (and thus quickly reaches a local minimum point as shown in Remark 1). Therefore, the heuristic step is important in order to allow further progress and push the coherence as low as possible in the maximum of $K$ iterations. As discussed in Remark 1 the mutual coherence is reduced, or kept at the same level, with every step solving \eqref{eq:IDCOMain3}. Clearly, the application of Theorem 2 of \cite{AlterningProjection} breaks the monotonicity and allows for an increase in coherence. Still, in many applications (for example \cite{GrassmannianApplication1}) where incoherent frames are desired there is an underlying assumption that the frame is also tight. In general, GFs are not tight frames.

Obviously, solving a sequence of such problems does not guarantee that we reach a maximally incoherent frame (GF or ETF) but considering that we actually solve a sequence of $\ell_\infty$ minimization problems we expect to have many components of the objective function equal in magnitude to the $\ell_\infty$ norm of the correlations between the current vector $\bm{f}_i$ with the others, and thus to be close to such a frame.

An extra advantage of this formulation is that, according to the design needs, it is very easy to add linear equality and convex inequality constraints. We use this fact in the results section to construct incoherent frames that also satisfy other convex constraints (like nonnegativity).

\section{Results}\label{sec:Results}

In this section we show the capabilities of SIDCO to: provide incoherence frames and sensing matrices for optimized compressed sensing. Finally, we also make a note on the creation of ETFs.

\subsection{Designing incoherent frames}

Given dimensions $(m,N)$ the first task is to use SIDCO in order to create frames as incoherent as possible.

Tables \ref{tb:IDCOn15}, \ref{tb:IDCOn25} and \ref{tb:IDCOm120} show the design capabilities of SIDCO for numerous frame dimensions, either for fixed dimension $m$ and increasing $N$ or vice-versa. In each case, we run SIDCO with $100$ different initializations. We show the mean mutual coherence of the initial frame, the minimum mutual coherence and the average mutual coherence reached by SIDCO in the $100$ runs. In each run, we compute the average mutual coherence defined as $\bar{\mu}(\bm{H}) = \frac{2}{N(N-1)} \sum_{i=1}^{N}\sum_{j=i+1}^N | \bm{h}_i^T \bm{h}_j |$ (not to be confused with the average coherence $\nu$ of \cite{AverageCoherence}). For perspective we also show the Welch bound (WB). The coherence levels reached are very close to this bound. In the initial tests, Tables \ref{tb:IDCOn15}, \ref{tb:IDCOn25} and \ref{tb:IDCOm120}, we design frames that are at least 2--overcomplete $(m, 2m)$ and maximally 10--overcomplete $(m, 10m)$. As expected, for larger $N$ the gap to the bound increases but is still kept in reasonable limits. Furthermore, notice that the mean mutual coherence reached is close to the minimum coherence, showing that the proposed method is not very sensitive to the choice of the initial frame. SIDCO runs for $K = 200$ iterations.
\def\arraystretch{1.05}
\begin{table}[t]
\begin{center}
\caption{Average incoherence results reached by SIDCO for final frames $\bm{H} \in \mathbb{R}^{15 \times N}$ with increasing values of $N$ for $100$ runs. The initial frames $\bm{H}_0$ are also shown together with the Welch bound (WB) for perspective.
}\label{tb:IDCOn15}
\begin{tabular}{|c|c<{\hspace{-5pt}}|c<{\hspace{-5pt}}|c<{\hspace{-5pt}}|c|c<{\hspace{-5pt}}|}
\hline
$N$ & avg($\mu(\bm{H}_0))$ & min($\mu(\bm{H}))$ & avg$(\mu(\bm{H}))$ & WB & avg$(\bar{\mu}(\bm{H}))$  \\ \hline
30	&	0.7264  &  0.2057  &  0.2073  & 0.1857 &   0.1841 \\ \hline
45    &	0.7429  &  0.2523  &  0.2602  & 0.2132 &   0.1941 \\ \hline
60    &	0.7554  &  0.2808  &  0.2866  & 0.2255 &   0.1994 \\ \hline
75    &	0.7654  &  0.3024  &  0.3072  & 0.2325 &   0.2028 \\ \hline
90    &	0.7736  &  0.3196  &  0.3230  & 0.2370 &   0.2051 \\ \hline
105  &	0.7806  &  0.3367  &  0.3395  &    0.2402 & 0.2067 \\ \hline
120	&	0.7867  &  0.3502  &  0.3585  &    0.2425 & 0.2077 \\ \hline
135   &	0.7922  &  0.3661  &  0.3715  &    0.2443 & 0.2083 \\ \hline
150   &	0.7969  &  0.3790  &  0.3830  &    0.2458 & 0.2086 \\ \hline
\end{tabular}
\end{center}
\end{table}
\begin{table}[t]
\begin{center}
\caption{Analogous to Table \ref{tb:IDCOn15} for final frames $\bm{H} \in \mathbb{R}^{25 \times N}$.
}\label{tb:IDCOn25}
\begin{tabular}{|c|c<{\hspace{-5pt}}|c<{\hspace{-5pt}}|c<{\hspace{-5pt}}|c|c<{\hspace{-5pt}}|}
\hline
$N$ & avg($\mu(\bm{H}_0))$ & min($\mu(\bm{H}))$ & avg$(\mu(\bm{H}))$ & WB  & avg$(\bar{\mu}(\bm{H}))$ \\ \hline
   50    &0.6343   & 0.1584  &  0.1599  &   0.1429 & 0.1420 \\ \hline
   75    &0.6491   & 0.1939   & 0.1954   &   0.1644 & 0.1499  \\ \hline
   100    &0.6600   & 0.2164  &  0.2182  &    0.1741 & 0.1541 \\ \hline
   120    &0.6937   & 0.2303   & 0.2378  &    0.1787 & 0.1568 \\ \hline
   125    &0.6690   & 0.2339    &0.2405  &    0.1796 & 0.1585 \\ \hline
   150    &0.6766   & 0.2476   & 0.2491  &    0.1832  & 0.1598 \\ \hline
   175    &0.6830   & 0.2589   & 0.2640  &    0.1857  &  0.1607 \\ \hline
   200    &0.6887   & 0.2695   & 0.2763  &    0.1876  & 0.1613\\ \hline
   225    &0.6937   & 0.2786   & 0.2799 &     0.1890  & 0.1618\\ \hline
   250    &0.6982   & 0.2874   & 0.2898 &    0.1901  & 0.1621\\ \hline
\end{tabular}
\end{center}
\end{table}
\begin{table}[t]
\begin{center}
\caption{Analogous to Table \ref{tb:IDCOn15} for final frames $\bm{H} \in \mathbb{R}^{m \times 120}$.
}\label{tb:IDCOm120}
\begin{tabular}{|c|c<{\hspace{-5pt}}|c<{\hspace{-5pt}}|c<{\hspace{-5pt}}|c|c<{\hspace{-5pt}}|}
\hline
$m$ & avg($\mu(\bm{H}_0))$ & min($\mu(\bm{H}))$ & avg$(\mu(\bm{H}))$ & WB & avg$(\bar{\mu}(\bm{H}))$ \\ \hline
   15  &0.8233   &0.3502    &0.3650  &0.2425 & 0.2138\\ \hline
   20  &0.7872   &0.2775    &0.2790  &0.2050 & 0.1995\\ \hline
   25  &0.6937   &0.2303    &0.2463  &0.1787 & 0.1879\\ \hline
   30  &0.6469    &0.1977    &0.1989  &0.1588 & 0.1780\\ \hline
   35  &0.6080    &0.1727    &0.1757  &0.1429 & 0.1695\\ \hline
   40  &0.5914    &0.1529    &0.1593  &0.1296 & 0.1620\\ \hline
   45  &0.5766    &0.1371    &0.1414  &0.1183 & 0.1552\\ \hline
   50  &0.5632    &0.1236    &0.1254  &0.1085 & 0.1491\\ \hline
   55  &0.5509    &0.1120    &0.1132  &0.0997 & 0.1435\\ \hline
   60  &0.5394    &0.1019    &0.1029  &0.0917 & 0.1382\\ \hline
\end{tabular}
\end{center}
\end{table}

The first comparison we make is between IDCO \cite{IDCO} and SIDCO. Table \ref{tb:IDCOn64} compares the two for the design of highly overcomplete frames (up to $m$-overcomplete) for dimension $m=64$. The IDCO results are lifted from \cite{IDCO}. We also show the Welch bound and $\lfloor s \rfloor$ from \eqref{eq:mcResult}, the maximum sparsity level of the signals that can be successfully recovered in each case by solving an $\ell_1$ minimization problem with the SIDCO frame. Both methods run for $K = 150$ iterations. Notice that the results of the two methods are close for small frames while the performance gap increases with $N$ in favor of SIDCO. In comparing these two methods, the big advantage of SIDCO is in terms of speed. For example, IDCO produces the $(64, 1280)$ frame in an over-week run while SIDCO terminates in an over-night run -- and actually with IDCO it is not computationally feasible to proceed for $m>1280$. The fact that each frame vector is updated individually means that SIDCO does not require large memory resources and the fact that the parameters $T_i$ are chosen to maximize performance (and not heuristically like for IDCO) explains the difference in performance. To show the versatility of the proposed method, SIDCO is used to produce nonnegative frames of the same dimensions $\bm{H}_{+} \in \mathbb{R}_{+}^{64 \times N}$ (all $h_{ij} \ge 0$). Because SIDCO deals directly with the frame (and not its Gram matrix) the nonnegativy constraints are trivial to add. In this particular setting we do not use step 2b) of SIDCO, that is not able in general to keep the entries nonnegative. We are not aware of another method that successfully designs frames with such a high degree of redundancy and/or nonnegative entries.

In compressed sensing, frames are usually called incoherent if the mutual coherence obeys $\mu(\bm{H}) \leq 1/\sqrt{m}$, the asymptotic coherence bound \eqref{eq:FrameBound} for large $N$ and the mutual coherence achieved by algebraically constructed ETFs. For all dimensions where examples are presented this bound is reached only for a small number of possible redundancies. In all the cases ($m=15$, $m=25$ and $m=64$) only the frames maximally 3--overcomplete obey this bound, with values $0.2582$, $0.2$ and $0.125$ respectively. The same holds for the results in Table \ref{tb:IDCOm120} where $N=120$ for $m \geq 40$.
\begin{table}[t]
\begin{center}
\caption{Incoherence results reached by IDCO and SIDCO for final frames $\bm{H} \in \mathbb{R}^{64 \times N}$ with increasing values of $N$. The initial frames $\bm{H}_0$ are also shown together with the Welch bound (WB) for perspective. The last column shows the coherence of the nonnegative frames $\bm{H}_{+} \in \mathbb{R}_{+}^{64 \times N}$ designed by SIDCO.
}\label{tb:IDCOn64}
\begin{tabular}{|c|c|c<{\hspace{-5pt}}|c<{\hspace{-5pt}}|c|c||c|}
\hline
$N$ & $\mu(\bm{H}_0)$ & $\mu(\bm{H})_\text{IDCO}$ & $\mu(\bm{H})_\text{SIDCO}$ &WB & $\lfloor \bar{s} \rfloor$ & $\mu(\bm{H}_+)$ \\ \hline
128	&0.3606	&0.0987	&0.0979	&0.0887 &5	&0.2480\\ \hline
192	&0.4596	&0.1206	&0.1186	&0.1023 &4	&0.2836\\ \hline
256	&0.4176	&0.1371	&0.1346	&0.1085 &4	&0.3058\\ \hline
320	&0.4755	&0.1451	&0.1429	&0.1120 &3	&0.3217\\ \hline
384	&0.5395	&0.1536	&0.1510	&0.1143 &3	&0.3369\\ \hline
448	&0.4993	&0.1599	&0.1580	&0.1159 &3	&0.3455\\ \hline
512	&0.5040	&0.1634	&0.1615	&0.1170 &3	&0.3539\\ \hline
576	&0.5009	&0.1720	&0.1693	&0.1180 &3	&0.3592\\ \hline
640	&0.5469	&0.1764	&0.1739	&0.1187 &3	&0.3675\\ \hline
768	&0.5250	&0.1894	&0.1854	&0.1198 &3	&0.3744\\ \hline
960	&0.6023	&0.2136	&0.1979	&0.1208 &3	&0.3878\\ \hline
1280	&0.6386	&0.2256	&0.2115	&0.1219 &2	&0.4042\\ \hline
1600	&0.6678	&--	&0.2230	&0.1225 &2	&0.4162\\ \hline
1920	&0.6890	&--	&0.2317	&0.1229 &2	&0.4272\\ \hline
2240	&0.6934	&--	&0.2394	&0.1232 &2	&0.4355\\ \hline
2880	&0.7001	&--	&0.2511	&0.1236 &2	&0.4476\\ \hline
3520	&0.7711	&--	&0.2604	&0.1239 &2	&0.4514 \\ \hline
4096	&0.8820	&--	&0.2617	&0.1240 &2	&0.4603 \\ \hline
\end{tabular}
\end{center}
\end{table}
\begin{figure}[t]
\centering
\includegraphics[width=0.45\textwidth]{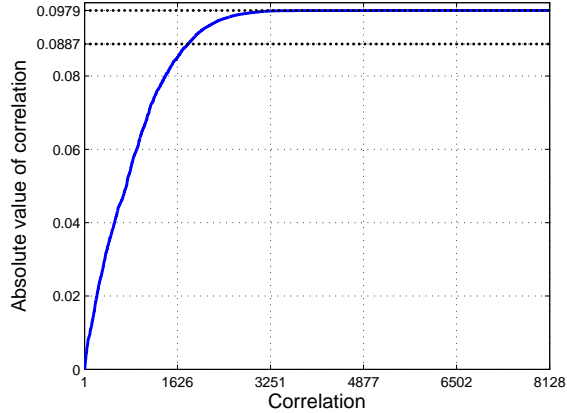}
\caption{The unique absolute value correlations of $\bm{H} \in \mathbb{R}^{64 \times 128}$. The figure shows their distribution and that many of them reach, or are close to, the mutual coherence value of the frame.}
\label{fig:Iterations64-128-correlations}
\end{figure}
\begin{figure}[t]
\centering
\includegraphics[width=0.45\textwidth]{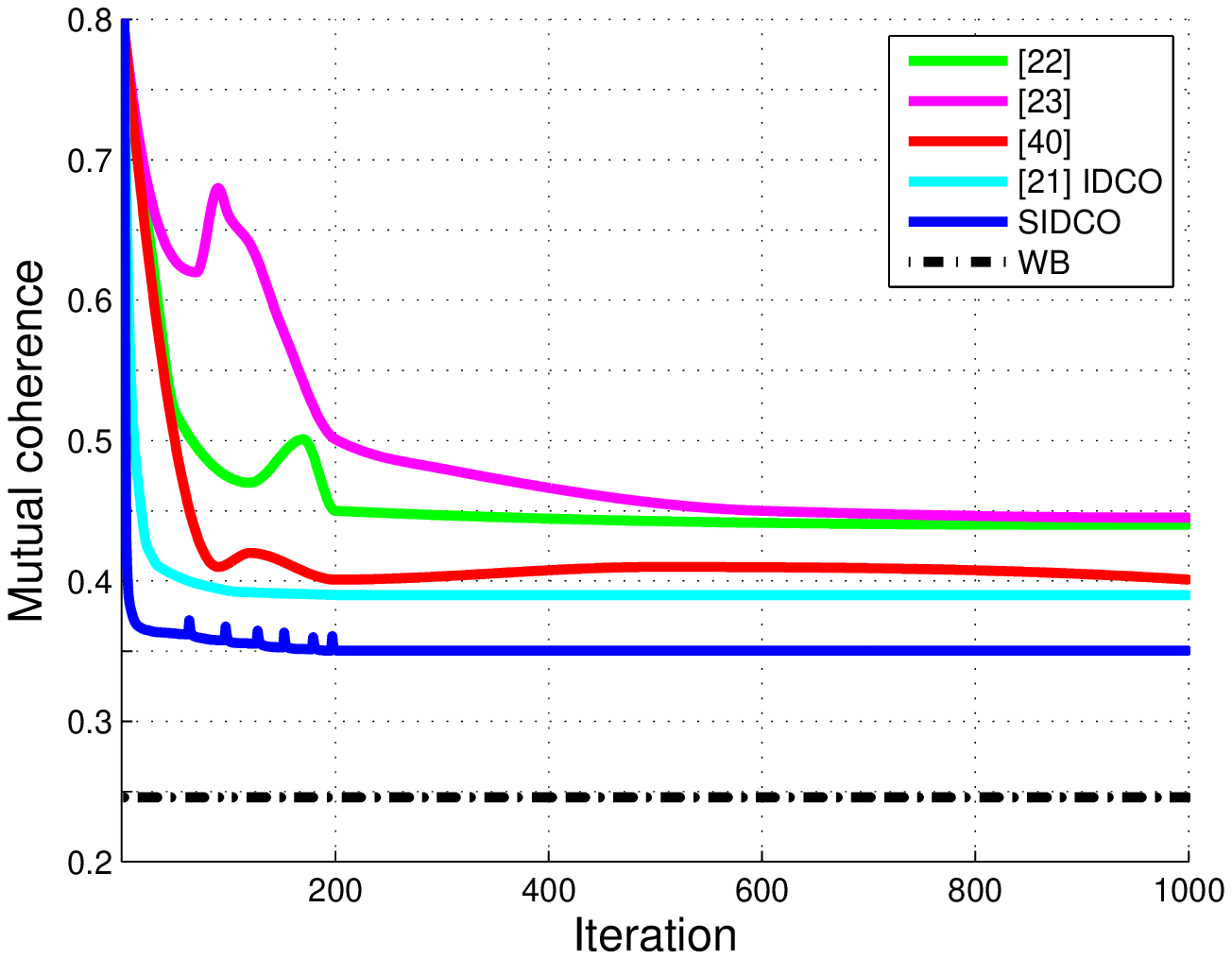}
\caption{Evolution by iteration of several algorithms for incoherent design of one frame $\bm{H} \in \mathbb{R}^{15 \times 120}$. The initial frames are the same for all algorithms.}
\label{fig:Iterations15x120}
\end{figure}
\begin{figure}[t]
\centering
\includegraphics[width=0.45\textwidth]{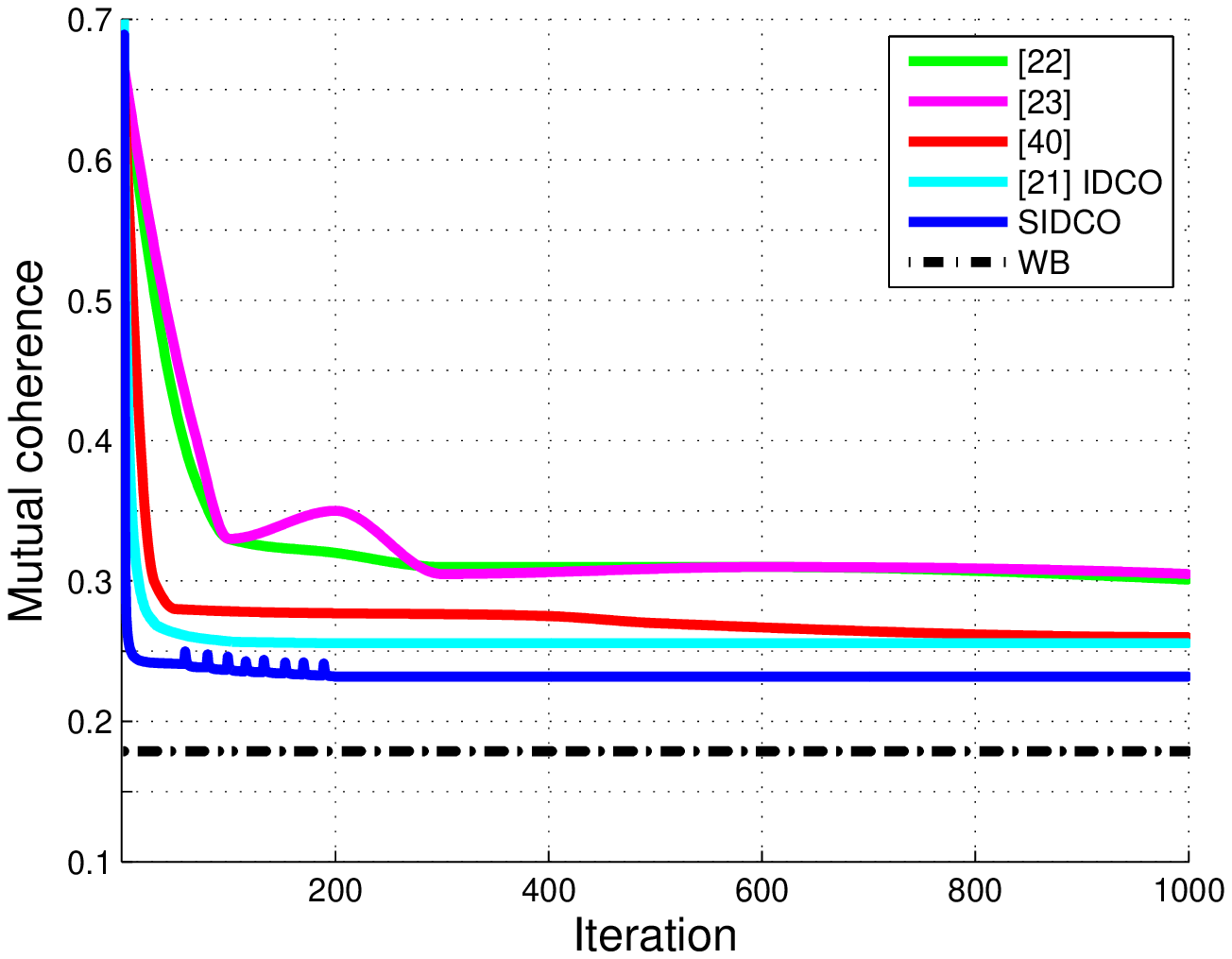}
\caption{Evolution by iteration of several algorithms for incoherent design of one frame $\bm{H} \in \mathbb{R}^{25 \times 120}$. The initial frames are the same for all algorithms.}
\label{fig:Iterations25x120}
\end{figure}
\begin{figure}[t]
\centering
\includegraphics[width=0.45\textwidth]{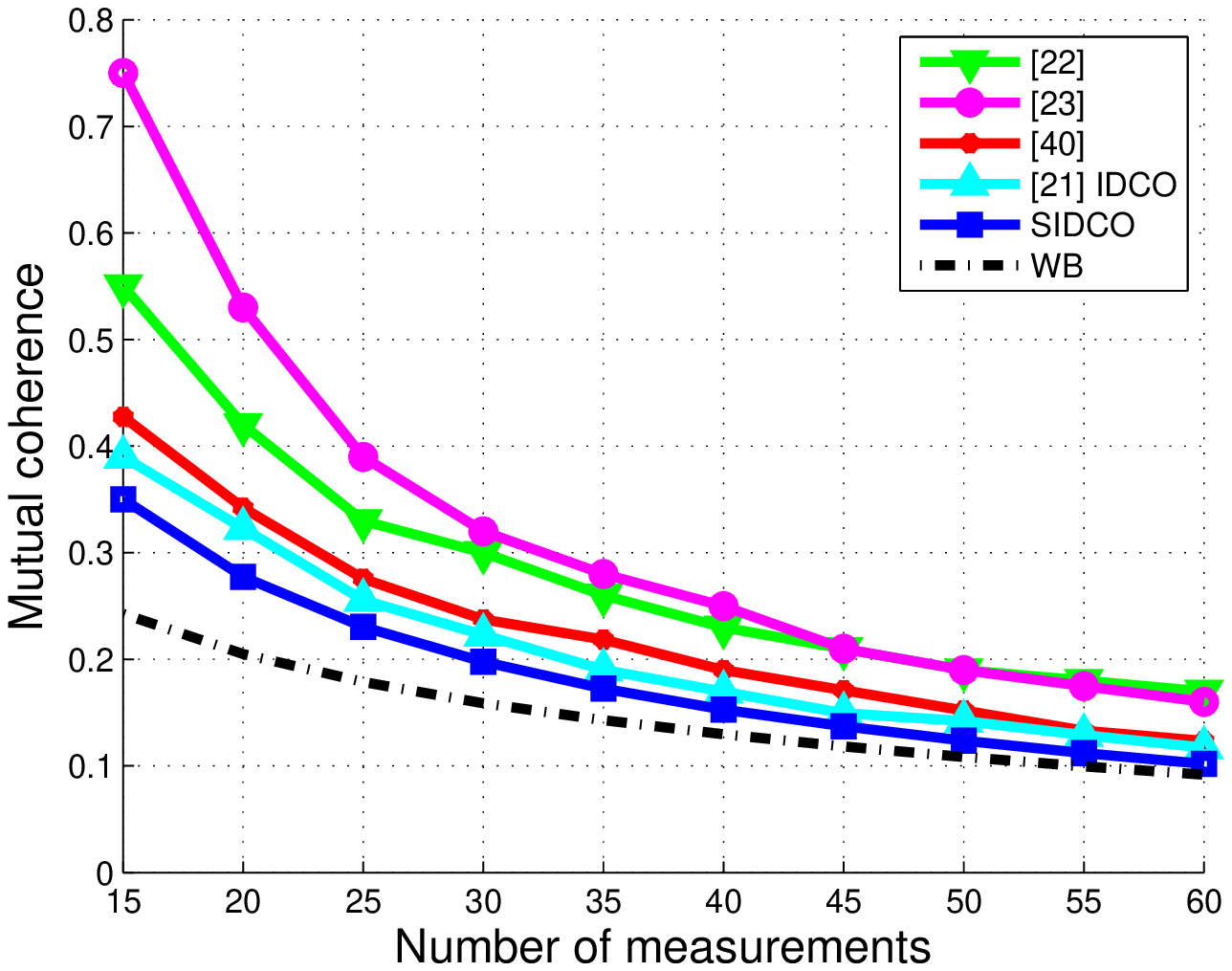}
\caption{Mutual coherence of several frames $\bm{H} \in \mathbb{R}^{m \times 120}$ designed by various algorithms. SIDCO is represented here by the minimum coherence column from Table \ref{tb:IDCOm120}. Results shown are averages obtained over 100 realizations. All algorithms are initializated with the same frames.}
\label{fig:n120}
\end{figure}

We now give some insights into the way SIDCO reaches highly incoherent frames and comparisons with competing methods. First, notice in Figure \ref{fig:Iterations64-128-correlations} the sorted unique correlations (in absolute value) between distinct vectors of the frame $\bm{H} \in \mathbb{R}^{64 \times 128}$ designed with SIDCO. The figure shows all 8128, $N(N-1)/2$ for $N=128$, unique correlations together with the mutual coherence reached $\mu(\bm{H})_\text{SIDCO}=0.0979$ and the Welch bound of 0.0887. We report that 66.14\% of the correlations are above $0.99\mu(\bm{H})_\text{SIDCO}$ and 74.21\% are above $0.95\mu(\bm{H})_\text{SIDCO}$. The figure shows how most of the correlations are close to the achieved mutual coherence.

Figure \ref{fig:Iterations15x120} shows the evolution, by iteration, of several algorithms for the design of frame $\bm{H} \in \mathbb{R}^{15 \times 120}$. SIDCO outperforms all methods in the literature. By the 5$^\text{th}$ iteration SIDCO achieves the mutual coherence that will be finally achieved by the third best approach \cite{Katsagellos}. The second best approach is IDCO. This plot highlights the difference between IDCO and SIDCO not just in terms of the final result but also, empirically, in terms of the rate of convergence. The difference originates in the way the parameters $T_i$ are chosen: heuristically for IDCO and by Remark 1 for SIDCO. This shows the effectiveness of the method. Also, for SIDCO the mutual coherence is decreasing at each iteration. Exceptions make a few iterations where spikes in the mutual coherence are recorded. These are the times when heuristic step 2b) is applied. The first time step 2b) is applied is at iteration number 62 where the mutual coherence is 0.3620. Until the 200$^\text{th}$ iteration step 2b) is applied 5 more times and the final mutual coherence reaches $\mu(\bm{H})_\text{SIDCO} = 0.3502$. Figure \ref{fig:Iterations25x120} is analogous to the previous figure but this time the frame in question is $\bm{H} \in \mathbb{R}^{25 \times 120}$. Again, the performance of \cite{Katsagellos} is reached in only 6 iterations while IDCO converges at a lower rate than SIDCO. In this situation the heuristic step 2b) is applied a total of 8 times: the first time SIDCO converges to a value of 0.2409 in 58 iterations and finally it reaches $\mu(\bm{H})_\text{SIDCO} = 0.2303$.

The SIDCO approach clearly outperforms all previous methods, in the first case quite significantly. In both these situations, SIDCO runs for 200 iterations and the other methods for 1000 iterations -- the result of SIDCO is extended to fit the plot. Both cases show that in the first iterations SIDCO makes significant progress while the later iterations provide additional progress but at a much lower pace. Running SIDCO for a large number of iterations will in general provide lower coherence levels. Taking this into account, in practice, if there is a time limit, SIDCO may run for only a few iterations. In these experimental runs we have used the generic convex optimization solver CVX \cite{CVX} -- it was not in the scope of this paper to build a custom solver. One may also argue that SIDCO needs to run only for a few iterations and that step 2b) can be ultimately avoided. This is due to the fact that SIDCO achieves low coherence very quickly, only after a few iterations, and then the progress due to step 2b) may be considered too little to be worth the large number of iterations that follow. Still, in a situation with no pressing time limit, SIDCO should run for a number of maximum iterations $K$ as large as possible.

\begin{figure}[t]
\centering
\includegraphics[width=0.46\textwidth]{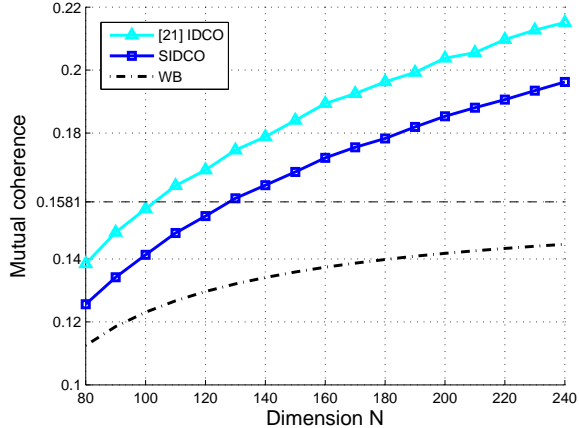}
\caption{Comparison between IDCO \cite{IDCO} and SIDCO for frames $\bm{H} \in \mathbb{R}^{40 \times N}$. We also highlight the $1/\sqrt{m} = 0.1581$ bound. Both algorithms are initialized with the same frames. SIDCO outperforms IDCO on average by 10\%.}
\label{fig:40m}
\end{figure}

Finally, Figures \ref{fig:n120} and \ref{fig:40m} compare the SIDCO method with the best competing methods in the literature and show again how it outperforms all of them. The difference in performance depends on the redundancy of the frames. In Figure \ref{fig:n120}, for highly redundant frames SIDCO outperforms the other methods by a significant margin (see $m \leq 25$) while otherwise the gap is diminished. Also, notice that for the frames with low redundancy (see $m \geq 35$) SIDCO's mutual coherence approaches the Welch bound so closely that no room for great improvement exists. In Figure \ref{fig:40m} we see how SIDCO fares against IDCO \cite{IDCO} for frames $\bm{H} \in \mathbb{R}^{40 \times N}$. We also show the $1/\sqrt{m} = 0.1581$ bound that is met again by frames at most 3--overcomplete. Since SIDCO is an improvement over IDCO there is no surprise that it outperforms it in all cases, the gap increasing with redundancy. In this situation, the heuristic step 2b) plays an important role in outperforming IDCO (to match the number of iterations).

For all frames we also compute the frame potential \eqref{eq:FP}. On average this value is within 1\% of its minimum value $(N^2/m)$.

\subsection{On the design of equiangular tight frames}

We also applied SIDCO to the design of real ETFs. Although SIDCO performs very well in all previous examples, and substantially outperforms previous methods, in this case it fails to achieve the exact structure of an ETF. We run SIDCO for the pairs $(m,N)$ with $N \leq 100$ for which we know real ETFs exist \cite{OnTheExistance} and it is virtually never able to exactly reach (with a coherence error of $10^{-5}$) the said ETF. This is even when in most cases the coherence is below the $1/\sqrt{m}$ bound. It seems these iterative approaches, without explicitly considering structural properties of ETFs (\cite{OnTheExistance, Holmes} and Remark 3), are not generally able to construct them. Note that we are never interested in the, trivial, real ETFs $(m,m)$ and $(m,m+1)$ that exist for every $m>1$ and are easily constructed \cite{OnTheExistance}. The difficulty of numerically constructing ETFs was also pointed out in \cite{AlterningProjection}.

\subsection{Optimized compressed sensing}
\begin{figure}[t]
\centering
\includegraphics[width=0.45\textwidth]{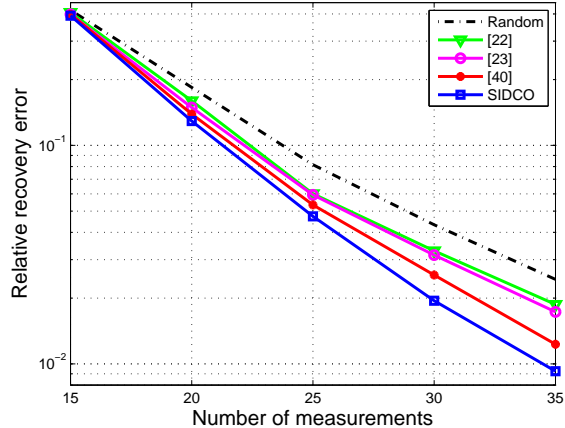}
\caption{Relative reconstruction errors for sensing matrices with $m$ measurements and fixed sparsity $s=4$ and fixed dimension $N=80$ for various algorithms. The reference performance is the random sensing matrix. For SIDCO the sensing frames have, in increasing order of $m$, the following coherences: 0.3067, 0.2416, 0.1980, 0.1690 and 0.1439.}
\label{fig:cs1}
\end{figure}
We now move to show the efficiency of the constructed frames for compressed sensing \cite{CS}. At the heart of compressed sensing we find the underdetermined system of linear equations $\bm{y} = \bm{Fx}$, where $\bm{F} \in \mathbb{R}^{m \times N}$ is called the \textit{sensing matrix} and where we know that $\bm{x}$ has a sparse representation like $\bm{x} = \bm{Da}$ where $\bm{a} \in \mathbb{R}^{M}$ is $s$-sparse in a known dictionary $\bm{D} \in \mathbb{R}^{N \times M}$. The goal is to recover $\bm{a}$ from a number of measurements $m$ as low as possible using a recovery algorithm. The correct recovery of $\bm{a}$ depends on the number of measurements $m$, the coherence of the sensing matrix $\bm{F}$ and the sparsity level of $\bm{a}$. We generate a random Gaussian dictionary $\bm{D}$ and random $s$-sparse representations $\bm{a}$ (with a random support) which we then try to recover using different sensing matrices (random or frames produces by various algorithms, including SIDCO). We show the results in synthetic experiments for $N = 80$, $M = 120$ and various values of $m$ and $s$. The sparse recovery is done using the Orthogonal Matching Pursuit (OMP) algorithm and the performance indicator is the relative energy content of the reconstructed signal $\bm{a}_\text{rec}$ as compared with the original: $\| \bm{a} - \bm{a}_\text{rec} \|_2^2 / \| \bm{a} \|_2^2$. For each choice of dimensions, we average the recovery results over $10^5$ runs are we present them when we vary the number of measurements $m$ for fixed sparsity $s = 4$ and fixed dimension $N = 80$. Results are shown in Figure \ref{fig:cs1}. Like in the previous section, SIDCO is better than every other method presented with the performance gap increasing for larger number of measurements.

As others point out \cite{Katsagellos}, the performance gain of using incoherent frames as sensing matrices does not scale as much as with the improvement in the mutual coherence. Observe that in Figure \ref{fig:cs1} for $m \in \{15, 20\}$ all results are very close to the random reference -- whereas the values of the mutual coherence of the initial, random, frames $\bm{H}_0$ in Tables \ref{tb:IDCOn15}, \ref{tb:IDCOn25}, \ref{tb:IDCOm120} and \ref{tb:IDCOn64} are very high compared to the mutual coherence achieved by SIDCO. To see why this might be the case we remind the reader that compressed sensing is able to recover the correct support when the amplitudes of the non-zero entries of $\bm{a}$ are above a constant times $\sqrt{M/m \log M}$ \cite{RandomRecovers}, which takes high values for relative small $m$ and that has been shown to hold for Gaussian random sensing matrices with order $s\log M$ measurements \cite{RandomRecovers}. Apart from the mutual coherence, recent developments \cite{CSMSE} show that unit norm tight frames perform well in compressive sensing applications when measuring the reconstruction average mean squared error.

\subsection{Incoherent dictionaries}

Observe that the construction of the incoherent frames is completely determined by the choice of dimensions $(m,N)$. These processes are blind to the nature of the data which will be represented in the frames. The idea in this section is to adapt the highly incoherent frames previously created to a given data sample such to reduce the reconstruction error.

We are presented with a sample dataset $\bm{Y} \in \mathbb{R}^{m \times M}$ and the goal is to adapt a given incoherent frame $\bm{F}_0$, designed by SIDCO, such that the reconstruction error of the provided dataset is minimized. We use the word "adapt" because we are interested only in transformations that preserve the incoherence of the frame $\bm{F}_0$. As such, we are interested in finding an orthogonal transformation $\bm{Q}$ such that $\bm{F} = \bm{QF}_0$ minimizes $\| \bm{Y} - \bm{FX} \|_F \| \bm{Y} \|_F^{-1}$ (the relative reconstruction error) where the sparse representations $\bm{X} \in \mathbb{R}^{N \times M}$ are found via some sparse approximation algorithm, like for example the Orthogonal Matching Pursuit (OMP) \cite{OMP}.

Consider the optimization procedure that at step $i$, given a fixed frame $\bm{F}_{i-1}$ and sparse representations $\bm{X}_{i-1}$, solves:
\begin{enumerate}
	\item Find $\bm{Q}_i$ via  the orthonormal Procrustes problem \cite{Proc}:
	\begin{equation}
		\underset{\bm{Q}_i; \ \bm{Q}_i \bm{Q}_i^T = \bm{Q}_i^T \bm{Q}_i = \bm{I}}{\text{minimize}} \ \  \|\bm{Y}-\bm{Q}_i \bm{F}_{i-1} \bm{X}_{i-1}\|_F^2, \\
		\label{eq:dictDesignNewFocused}
	\end{equation}
with solution $\bm{Q}_i = \bm{UV}^T$ from $\bm{Y} \bm{X}_{i-1}^T \bm{F}_{i-1}^T = \bm{U\Sigma V}^T$.

	\item Update the dictionary $\bm{F}_i = \bm{Q}_i \bm{F}_{i-1}$.

	\item Construct the new representations $\bm{X}_i$ using $\bm{F}_i$ and the OMP algorithm with target sparsity $s$ on the dataset $\bm{Y}$.
\end{enumerate}
The final solution is the accumulation of the rotations, say after $K$ steps, into $\bm{F}_K = \bm{Q}_K \dotsb \bm{Q}_2\bm{Q}_1\bm{F}_0$ which preserves mutual coherence, i.e., $\mu(\bm{F}_K) = \mu(\bm{F}_0)$. For simplicity, herein we denote $\bm{F}=\bm{F}_K$.
\begin{figure}[t]
\centering
\includegraphics[width=0.45\textwidth]{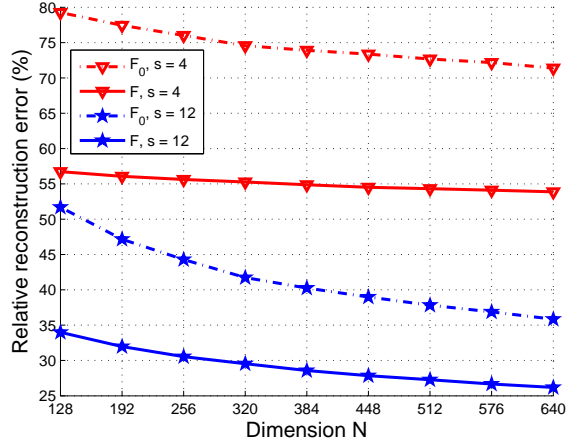}
\caption{Relative reconstruction errors of several dictionaries $\bm{F} \in \mathbb{R}^{64 \times N}$ using the training procedure described in this section starting from frames $\bm{F}_0$ previously designed by SIDCO. We consider two sparsity levels $s \in \{4, 12 \}$. We also show in dotted lines the reconstruction errors for initial frames $\bm{F}_0$.}
\label{fig:RecError}
\end{figure}
This optimization procedure was used in \cite{Rotations} together with an iterative projections approach to design incoherent overcomplete dictionaries and in \cite{Initialization} as the basis for an initialization strategy for general dictionary learning. In this section we propose to use the highly incoherent frames created via SIDCO and then use the fast rotational transformations to adapt it to a given dataset. This way, the coherence of the dictionary is fixed during the whole training procedure while the reconstruction error is reduced.

We use image data from popular test images (Lena, peppers, etc.) by extracting non-overlapping blocks of $8\times8$ pixels from which we remove the means and we subsequently normalize. The overall training matrix is $\bm{Y} \in \mathbb{R}^{64 \times 16384}$. As initial frames we use the incoherent frames $\bm{F} \in \mathbb{R}^{64 \times N}$ previously created and run the training procedure based on rotations for $K=500$ iterations. We show in Figure \ref{fig:RecError} the relative construction errors achieved by the incoherent frames before, $\bm{F}_0$, and after training, $\bm{F}$. The improvement is obvious but notice that it does not scale with dimension $N$ due to the incoherent structure.

\section{Conclusions}\label{sec:Conclusions}

In this paper we present a method for the design of highly incoherent frames that operates directly on the frame vectors and is based on solving a series of convex optimization problems that approximate the original nonconvex design problem. We derive conditions under which the algorithm converges and show that equiangular tight frames are in fact local minimum of this method. Extended simulations, covering low and high dimensional spaces and different levels of redundancy, show that the proposed method substantially outperforms all previously proposed incoherent frame design methods.


\end{document}